\documentstyle[preprint,aps]{revtex}
\input{epsf}

\begin{document}
\draft
\title{Localization in Fock space: A finite size scaling hypothesis
for many particle excitation statistics}

\author
{Richard Berkovits and Yshai Avishai$^{\dag}$}

\address{
The Minerva Center for the Physics of Mesoscopics, Fractals and Neural 
Networks,\\ Department of Physics, Bar-Ilan University,
Ramat-Gan 52900, Israel}

\address{
$\dag$also at Department of Physics, Ben-Gurion University, Beer-Sheva, Israel}

\date{\today}
\maketitle

\begin{abstract}
The concept of
localization in Fock space is extended to the study of the
many particle excitation statistics
of interacting electrons in a two dimensional quantum dot.
In addition, a finite size scaling hypothesis for
Fock space localization, in which
the excitation energy replaces the system size,
is developed and tested by analyzing
the spectral properties of the quantum dot.
This scaling hypothesis, modeled after the usual Anderson
transition scaling, fits the numerical data obtained for
the interacting states in the dot.
It therefore attests to the relevance of
the Fock space localization scenario
to the description of many particle excitation properties.

\end{abstract}


\narrowtext

\newpage

The concept of localization in Fock space has been recently utilized\cite{agkl}
to explain the transition in the width of excited states measured
in tunneling conductance experiments for quantum dots\cite{siv}. 
Due to electron-electron interactions the quasi-particle states are coupled,
which leads to their finite width. At a critical value of interaction
strength and excitation energy the width becomes essentially
infinite and the quasi-particle states can no longer be resolved.

In this letter we show that 
Localization in Fock space explains 
also the recently reported\cite{ber,ba,js,wpi}
transition in
the level statistics of excited many particle
states as a function of the interaction
strength. For small
values of the interaction the many-particle states are localized in the 
non-interacting Fock space, namely each interacting state is
composed out of a small number of 
non-interacting eigenvectors. Above some critical value they are extended, 
i.e., composed out of a large number of non-interacting eigenstates.
This leads to a transition in the statistical properties of the energies
from Poisson to Wigner statistics, which as in Anderson localization
is a true second order phase transition. We will demonstrate this by 
constructing a finite-energy scaling theory
which plays the role of finite size scaling familiar in standard
localization theory.

\begin{figure}
\centerline{\epsfxsize = 4in \epsffile{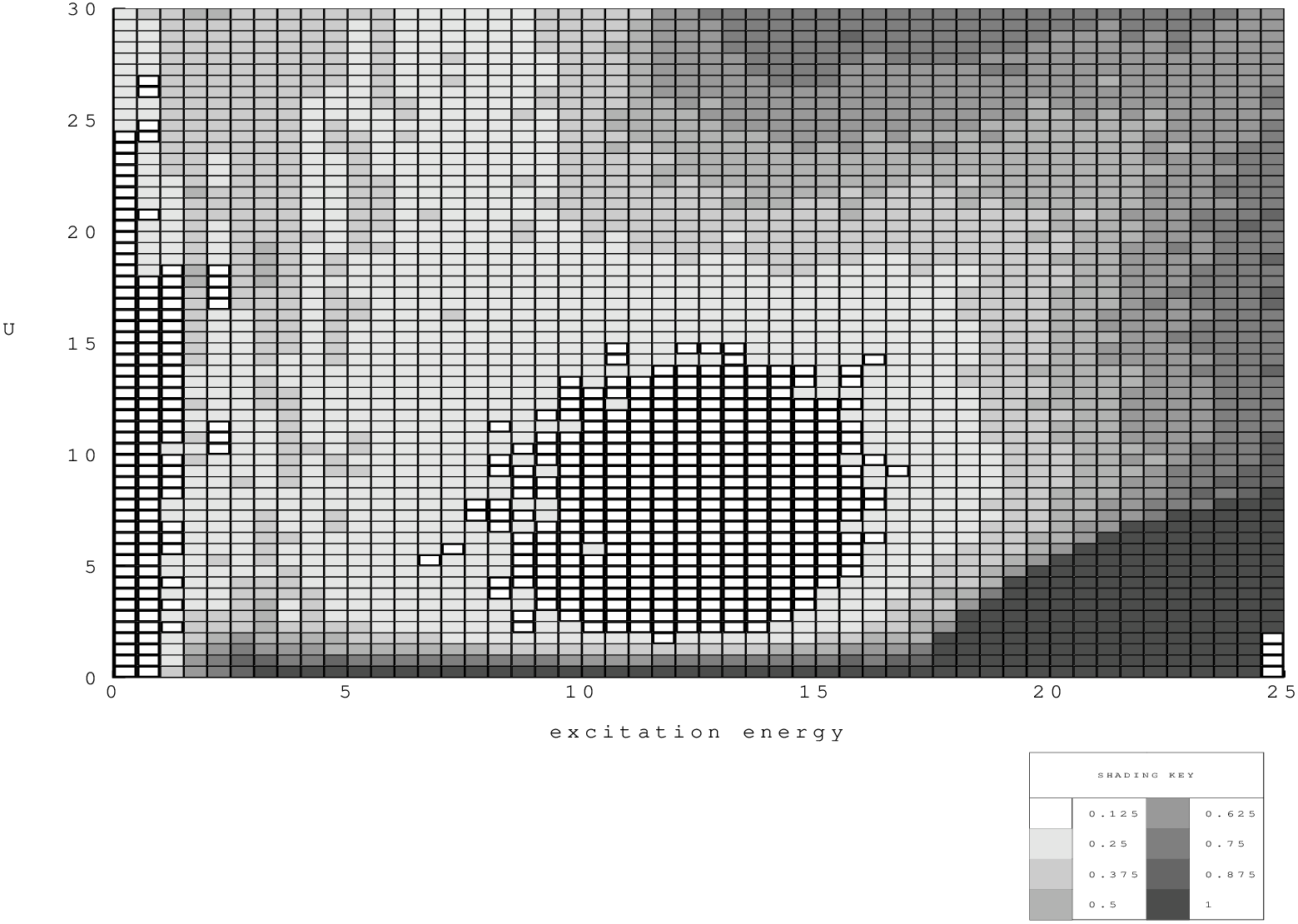}}
\caption{A gray-scale map of the parameter $\gamma$ defined in
Eq. (\protect \ref{Gamma}) for the many particle Hamiltonian
given in Eq. (\protect \ref{hamil0}). Low values of $\gamma$ correspond to
Wigner statistics, while values close to one correspond to
Poisson statistics }
\label{fig1}
\end{figure}

A typical gray-scale map of the statistical properties in the space
of interaction strength and excitation energy
is presented in Fig. \ref{fig1}.
at low excitation energies the statistics remains Wigner over a large range of
interaction strength. This will be discussed in detail elsewhere. A 
transition from Poisson to Wigner and back to Poisson as the interaction
strength is increased is evident for intermediate
excitation energies.
The second (Wigner-Poisson) transition is the result of Wigner
crystallization \cite{ba}. In this letter we will show that the 
first transition has the characteristics of a localization transition in
Fock space.

A localization transition is characterized by a single parameter scaling
behavior\cite{aalr}, which results in only three possible types
of statistics in the thermodynamic limit: Poisson in the localized regime,
Wigner in the extended one and a possible third type of statistics
at the transition point\cite{sssls,klaa}. 
As shown by Shklovskii et. al. \cite{sssls}, 
for a finite system the transition between
the two types of statistics is gradual and may be characterized by
a finite size scaling behavior, which can be used to locate the critical point
and the critical exponent.

For the Anderson localization transition the scaling parameter
is the ratio of the disorder dependent localization length to the linear
dimension of the system. In the Fock space localization transition we 
show that the relevant scaling variable is the ratio 
of the localization length
(which depends on the interaction strength and the two-particle density) to
the linear ``dimension'' of the system in Fock space,
which depends mainly  on the excitation
energy. The latter is surprising, since one would expect 
the linear dimension of the system to depend on the number of particles
and the number of sites. As has been demonstrated in
Ref. \cite{js}, no clear finite size scaling behavior is observed as function
of the number of sites or particles. This is the result of the
unusual ``geometry'' of the interaction coupling which leads to
a Cayley tree structure in Fock space where the number of generations
depends on the excitation energy\cite{agkl}. 

Let us begin by defining the Fock space. 
The many-electron second quantization Hamiltonian:
\begin{eqnarray}
H=H_0+H_{int}=\sum_{i} \epsilon_i  c_i^{\dag}  c_i +
\sum_{i,j,k,l} U_{i,j,k,l} c_j^{\dag}  c_l^{\dag}  c_k c_i,
\label{hamil}
\end{eqnarray}
where $c_i^{\dag}$ is the creation operator of particle in the i-th single 
electron state, $\epsilon_i$ is its energy, and $U_{i,j,k,l}$ is the
interaction matrix element,
has a set of eigenvectors $|\Psi_N^j\rangle$ and eigenvalues 
$E_N^j$,
where $N$ is the number of electrons and $j$ is an index ordering the
states by ascending eigenvalues.
There are $M=(_N^S)$ many-electron states,
where $S$ is the number of single electron states.

For the non-interacting ($H_{int}=0$) case the ground-state
eigenvector and eigenvalue are
\begin{eqnarray}
|\Phi_N^0\rangle = c_N^{\dag} \ldots c_1^{\dag} |0\rangle
\nonumber\\
E_N^0=\sum_{i=1}^N \epsilon_i,
\label{nintgs}
\end{eqnarray}
here $|0\rangle$ is
the vacuum state. An excitation composed of $m$ electron-holes pairs 
may be represented as:
\begin{eqnarray}
|\Phi_N^{\alpha_1 \ldots \alpha_{2m}}\rangle = 
c_{\alpha_{2m}}^{\dag} \ldots c_{\alpha_{m+1}}^{\dag} c_{\alpha_{m}} 
\ldots c_{\alpha_{1}} |\Psi_N^0\rangle
\nonumber\\
E_N^{\alpha_1 \ldots \alpha_{2m}}=E_N^0 - 
\sum_{i=1}^m \epsilon_{\alpha_{i}} + \sum_{i=m+1}^{2m} \epsilon_{\alpha_{i}}.
\label{nintex}
\end{eqnarray}
These excitations may also be numbered in ascending order $|\Phi_N^j\rangle$ 
according to their energy.

For the interacting ($H_{int}\ne0$) case one may write the eigenvectors 
$|\Psi_N^j\rangle$ and
their excitation energy $E_N^j -E_N^0$ (again arranged in ascending order)
in the following way:
\begin{eqnarray}
|\Psi_N^j\rangle = \sum_{m=1}^N \sum_{\alpha_1 \ldots \alpha_{2m}}
A^{\alpha_1 \ldots \alpha_{2m}} |\Phi_N^{\alpha_1 \ldots \alpha_{2m}}\rangle
= \sum_{k=1}^M A_k |\Phi_N^k\rangle,
\nonumber\\
\varepsilon_j=E_N^j -E_N^0.
\label{intex}
\end{eqnarray}
An interesting question is how are these interacting eigenstates composed out
of the non-interacting Fock space eigenvectors. It is natural to expect that
for small values of interaction and excitation energy an interacting eigenstate
$|\Psi_N^j\rangle$ will be composed of a small number of non-interacting
eigenvectors $|\Phi_N^k\rangle$ where $k$ is in the vicinity of $j$, while
for large values of the interaction $|\Psi_N^j\rangle$  will be composed of
many non-interacting eigenvectors. Altshuler et. al. \cite{agkl} have
termed this transition ``localization in Fock space'' and predicted that
it should exhibit the characteristics of an Anderson transition on a
Cayley tree.

We will use two measures for the Anderson transition: (i) the
energy level statistics, and (ii) the inverse participation ratio
in Fock space. A convenient way to characterize the change
in the statistics of a system proposed
by Shklovskii et. al. \cite{sssls} is to study the parameter $\gamma$
defined as
\begin{equation}
\gamma= {{\int_2^\infty P(s) ds - e^{-\pi}}\over{e^{-2}- e^{-\pi}}},
\label{Gamma}
\end{equation}
where $P(s)$ is the distribution of the normalized level spacings 
$s=E_N^j -E_N^{j-1}/\langle E_N^j -E_N^{j-1} \rangle$,
where $\langle \ldots \rangle$ denotes an average over different
realizations of disorder. For an infinite system $\gamma$
changes sharply from $\gamma=1$ in the localized regime to $\gamma=0$ 
in the extended regime. For a finite system
the change is gradual. One can then use a finite size scaling 
hypothesis to identify
the transition point and its critical indices\cite{sssls}. 
The inverse participation ratio $P$, defined as
\begin{eqnarray}
P=\sum_{k=1}^M |A_k|^4,
\label{par}
\end{eqnarray}
is expected to change in the Anderson transition for an infinite system 
from $P$ which is a function of the localization length
in the localized regime to
$P=0$ in the extended regime. 

The argument for a localization transition as one increases the
interaction strength follows the usual Anderson transition
argument\cite{at}. The strength of the
hopping term $V$ coupling neighboring sites is compared with
the inverse of the 
density of states of the neighboring site, $K \nu$, where
$\nu=1/W$, $W$ being the strength of disorder, and K is the connectivity. 
A localization transition will occur at  a critical value
\begin{eqnarray}
Z_c = K \nu V = {{K V}\over{W}} \sim 1.
\label{crit}
\end{eqnarray}
As noted by
Altshuler et. al. \cite{agkl}, a similar argument may be applied to
Fock space localization. The two-body interaction couples
non-interacting eigenvectors which are different by up to 
2 electron-hole pairs. Assuming that the main coupling is between
states differing by a single electron-hole pair\cite{agkl},
the average strength of coupling
\begin{eqnarray}
\tilde U = \langle \langle \Phi_N^{\alpha_1' \ldots \alpha_{2m}'}|
H_{int}|\Phi_N^{\alpha_1 \ldots \alpha_{2m}}\rangle 
\rangle_{\alpha_1',\alpha_{2m}'},
\label{ut}
\end{eqnarray}
where $\langle  \ldots 
\rangle_{\alpha_1',\alpha_{2m}'}$ denotes averaging over all
possible electron-hole pairs. The density of such coupled states at
excitation energy $\varepsilon$ is denoted by $K \nu_2(\varepsilon)$. Thus
\begin{eqnarray}
Z_c \sim K \nu_2(\varepsilon) \tilde U.
\label{crit1}
\end{eqnarray}
This argument is similar to the one used by Imry in the context of two 
particle state delocalization\cite{im,wpi}.
A more careful consideration of the geometry of the connected states
reveals that it is similar to a Cayley tree for which
$K$ should be replaced by 
$K \ln K$ \cite{aat}. 
Since the effective connectivity $K \sim g$ where $g$ is the dimensionless
conductance, this is a correction of order of $\ln g$.
Altshuler et. al. \cite{agkl} concluded that due to the Cayley tree structure,
an intermediate region $Z_c/K > \nu_2(\varepsilon) \tilde U > Z_c/K \ln K$
between the localized region $\nu_2(\varepsilon) \tilde U < Z_c/K \ln K$
and the extended region $Z_c/K < \nu_2(\varepsilon) \tilde U $ should exist,
which is non-ergodic. Using a non-linear sigma model for the Cayley tree
Mirlin and Fyodorov found no intermediate region\cite{fm}.

Here we would like to extend this Fock space localization 
picture to include one of the most useful concepts in the Anderson
localization picture, namely that of finite size scaling. In the usual
Anderson picture $\gamma(K V/W,L) = f[L/\xi(K V/W)]$\cite{sssls} where $L$ 
is the sample size, $\xi(K V/W)$ is the localization length and $f$ is a
scaling function. Near the critical value it is expected to behave
as $\gamma(K V/W,L) = \gamma(Z_c,L) + C [Z_c - (K V/W)] L^{1/\delta}$,
where  $C$ is a constant and $\delta$ is the critical exponent
(usually denoted by $\nu$).
From the above discussion it is natural to assume that 
$ K \nu_2(\varepsilon) \tilde U$ will replace $K V/W$ for the Fock
space localization. The size of the system, i.e., the number of generations
in the Cayley tree, is equivalent to the number of electron-hole
pairs which can be generated at a given excitation energy\cite{agkl}. Thus
$L \sim n_{max} \sim \sqrt{\varepsilon/\Delta}$, where $\Delta$ is the single
electron level spacing. Hence we expect
\begin{eqnarray}
\gamma(K \nu_2(\varepsilon) \tilde U, \varepsilon) \sim
\gamma(Z_c,\varepsilon) + 
C \left(Z_c - K \nu_2(\varepsilon) \tilde U \right) 
\varepsilon^{1/2\delta}.
\label{scal}
\end{eqnarray}

For
$P(K \nu_2(\varepsilon) \tilde U, \varepsilon)$, the situation is 
different. According to the Breit-Wigner formula\cite{fic,fcic}
in the localized regime the inverse participation ratio depends on 
$U$ and $K \nu_2(\varepsilon)$ but not on the linear dimension $\varepsilon$,
as long as $\xi(K \nu_2(\varepsilon) \tilde U)<L$.
Once $\xi(K \nu_2(\varepsilon) \tilde U)>L$, $P$ will also depend on
$\varepsilon$. Since $\xi\propto (Z-Z_c)^{-\delta}$ we expect that $P$
for different values of $\varepsilon$ to coalesce for $Z \ll Z_c$
and to begin to fan out for $Z \sim Z_c$.

In order to check this finite size
scaling hypothesis we have numerically calculated
$\gamma$ and $P$ for a specific many particle tight-binding Hamiltonian:
\begin{eqnarray}
H= \sum_{k,j} \omega_{k,j} a_{k,j}^{\dag} a_{k,j} - V \sum_{k,j}
(a_{k,j+1}^{\dag} a_{k,j} + a_{k+1,j}^{\dag} a_{k,j}) + h.c
+ H_{int},
\label{hamil0}
\end{eqnarray}
where  
$\omega_{k,j}$ is the energy of a site ($k,j$), chosen 
randomly between $-W/2$ and $W/2$ with uniform probability, and $V$
is a constant hopping matrix element.
The interaction Hamiltonian is given by:
\begin{equation}
H_{int} = U  
\sum_{k,j>l,p} {{a_{k,j}^{\dag} a_{k,j}
a_{l,p}^{\dag} a_{l,p}} \over 
{|\vec r_{k,j} - \vec r_{l,p}|/b}}
\label{hamil2}
\end{equation}
where $U=e^2/b$ and
$b$ is the lattice unit. 

We consider a $4 \times 3$ dot with $S=12$ sites and $N=4$ 
electrons. The $M \times M$ Hamiltonian
matrix is numerically
diagonalized and all the eigenvectors $|\Psi_N^j\rangle$ and eigenvalues
$E_N^J$ are obtained. 
The strength $U$ of the interaction is varied between $0-30V$. 
The disorder strength is chosen to be $W=5V$
in order to assure metallic behavior
for the non-interacting case.
This is important
in order to obtain a reasonable connectivity $K\sim g$. 
For each value of $U$, the results are averaged over
$1000$ different realizations of disorder.
$\gamma(U,\varepsilon)$ is directly calculated from the level spacing 
distribution for different values of $U$ and $\varepsilon$,
while $P(U,\varepsilon)$ is calculated from the overlap of
$|\Psi_N^j\rangle$ with the non-interacting eigenvectors.

\begin{figure}
\centerline{\epsfxsize = 4in \epsffile{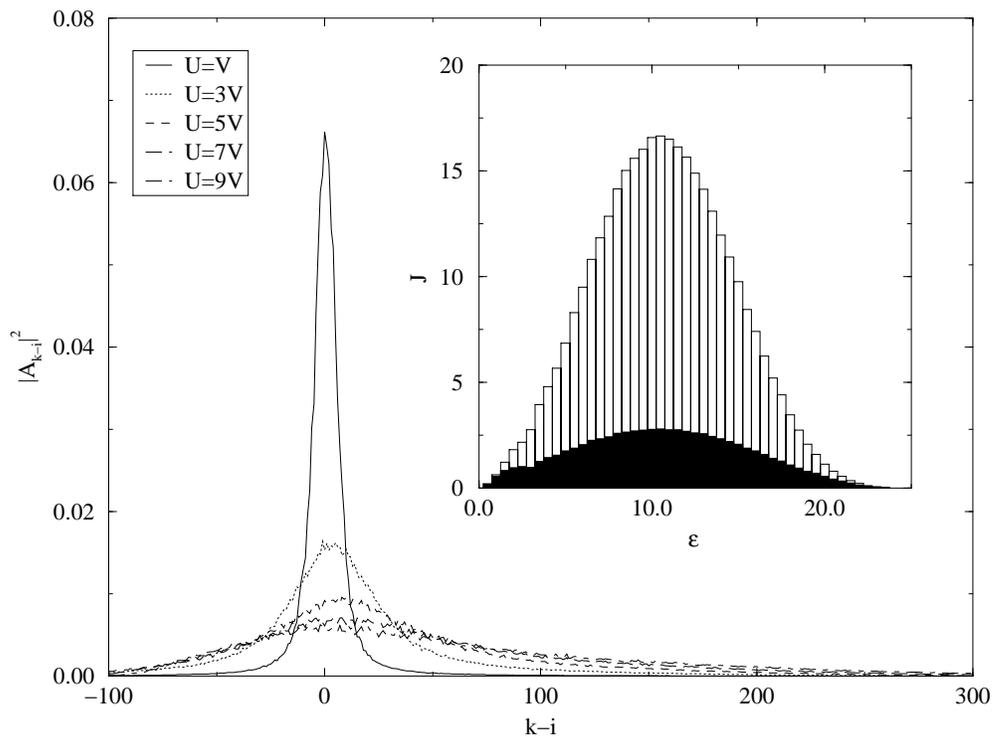}}
\caption{The non-interacting Fock space composition of an interacting
eigenvalue of excitation energy $\protect\varepsilon\sim 8V$. The composition
$|A_{k-i}|^2$ was averaged over 100 realizations of disorder. Inset: the number
of states coupled by a single pair transition $J$ within an energy bin
(dark histogram) compared with the total number of states (white histogram).}
\label{fig2}
\end{figure}

In Fig. \ref{fig2} the average value of $|A_{k-i}|^2$ for an excited state
$|\Psi_N^i\rangle$ (where $i$ is chosen so that $\varepsilon_i \sim 8V$ in the 
non-interacting case)
as a function of the interaction strength is plotted. As expected 
for small values of interaction $|A_{k-i}|^2$ is strongly peaked around
$k=i$. Thus, the interacting state is composed of a small number of
non-interacting states of similar energies, which has on the average a
Breit-Wigner form\cite{fic,fcic}.
As the interaction increases the interacting state is composed of
an increasing number of non-interacting states although there is no obvious
sharp transition. Similar features appear in $|A_{k-i}|^2$ for a single
realization, although the curves are noisier.

In order to locate the transition point we use finite size scaling. First
we calculate $K \nu_2(\varepsilon) \sim J(\varepsilon)$ by evaluating the 
number $J(\varepsilon)$ of
non-interacting states in a given energy region which are coupled to each 
other by a single electron-hole transition. The numerically calculate  
$J(\varepsilon)$ is plotted in
the inset in Fig. \ref{fig2}. For small values of $\varepsilon$, $J$ increases
linearly, while near the middle of the excitation band it flattens out.
Since the interaction strength $\tilde U \propto U$ one may replace
the scaling parameter
$K \nu_2(\varepsilon) \tilde U$ by  $J(\varepsilon) U$.

\begin{figure}
\centerline{\epsfxsize = 4in \epsffile{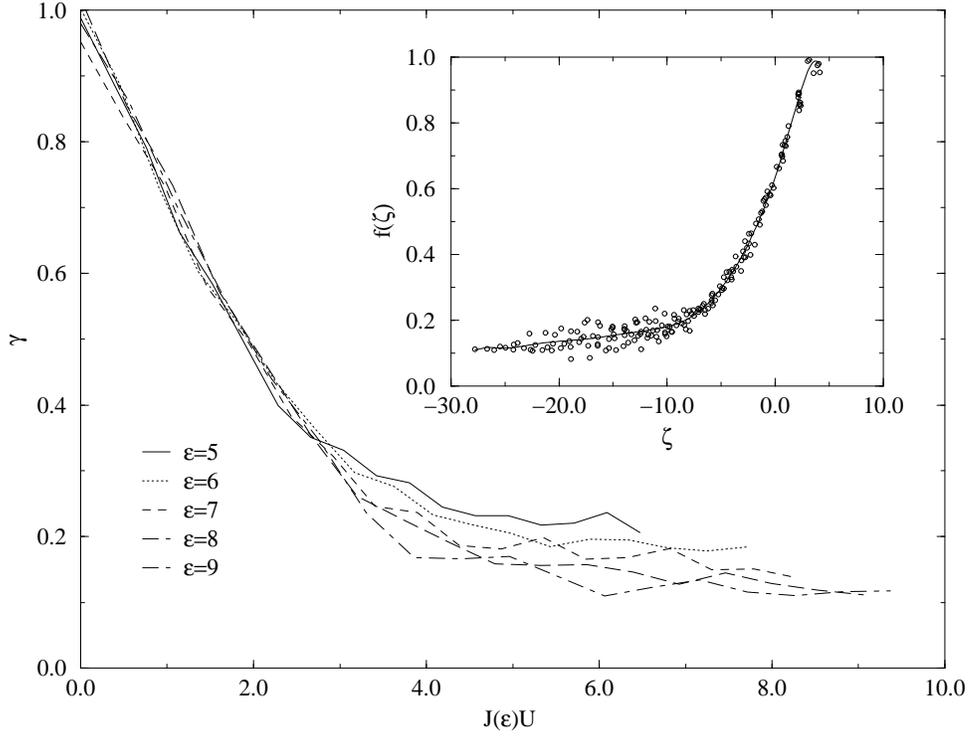}}
\caption{The parameter $\gamma$ for different values of interaction $U$
and excitation energy $\varepsilon$. Inset: the scaling function
$f(\zeta)$.}
\label{fig3}
\end{figure}

The parameter $\gamma(J(\varepsilon) U, \varepsilon)$ 
is shown in Fig. \ref{fig3} for values of $5V<\varepsilon<10V$.
This range was chosen in order to avoid the special region of low excitations
and to avoid the symmetric region above the middle of the band 
(see Fig. \ref{fig1}).
Indications of finite size scaling behavior are observed. For large values
of $J(\varepsilon) U$, $\gamma$ becomes smaller as the size $\varepsilon$
increases. On the other hand for small values of $J(\varepsilon) U$, 
$\gamma$ tends to be larger as the size increases. The point at which the lines
intersect is the transition point. 
The quantity $\gamma(J(\varepsilon) U, \varepsilon)$ is then 
considered as a
certain scaling function $f(\zeta)$ of the scaling variable
$\zeta=\varepsilon^{1/2 \delta} (Z_c-J(\varepsilon) U)$. 
Practically, it is useful to shift the
variable $\zeta$ to $y(\zeta)=\frac {\zeta-a-b} {b-a}$ where $a$ and $b$ are
respectively the minimum and maximum values assumed by $\zeta$.
Evidently, $y(\zeta)$ ranges between $-1$ and $1$. 
Then one expands $f(\zeta)$ in a series
of Chebyshev polynomials $T_{n}[y(\zeta)]$
($n=0,1,2,...Q$). 
Minimization of the set of differences 
$|f(\zeta)-\gamma(J(\varepsilon) U, \varepsilon)|$ results in the unknowns
$Z_c$, $\delta$ and the expansion coefficients (namely, the
scaling function itself). In all cases, it is sufficient
to cut off the number of polynomials at $Q=10$.
The best fit is obtained for $Z_c=1.3\pm0.4$ and $\delta=1\pm0.3$.
The function $f(\zeta)$ and the original data points are plotted in the 
inset of Fig. \ref{fig3}. 
 
\begin{figure}
\centerline{\epsfxsize = 4in \epsffile{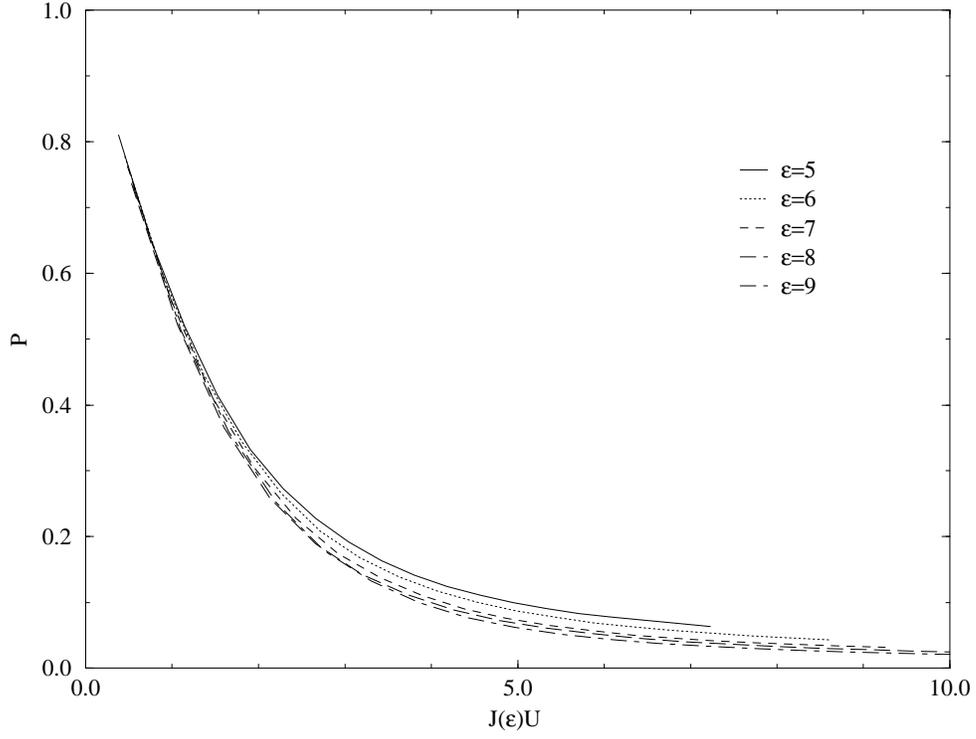}}
\caption{The inverse participation ratio $P$
or different values of interaction $U$
and excitation energy $\varepsilon$. }
\label{fig4}
\end{figure}

The results for the inverse participation ratio $P$
are presented in Fig. \ref{fig4}. For $J(\varepsilon) U<Z_c$
all lines coalesce, since no dependence on the linear dimension
$\varepsilon$ is expected. 
For $J(\varepsilon) U \sim Z_c$ the dependence on $\varepsilon$ is apparent
and the lines corresponding to the different values of $\varepsilon$
fan out.
The region in which the lines fan out 
correspond to $Z_c\sim1.5$, which is close to the
values obtained for $\gamma$. It is interesting to note that
$\gamma(Z_c)\sim 0.6$ is considerably higher than for 
the Anderson transition\cite{sssls}. 

In this analysis we have not clearly identified a crossover region
between the localized and extended regions predicted in Ref. \cite{agkl}.
This may be explained by the results of Ref. \cite{fm}, although
we do not see the jump in $P$ predicted there.
We speculate that the additional couplings between generations which are
not taken into account
in the pure Cayley tree picture tend to restore ergodicity, which is a 
key ingredient in the scaling picture.
Only studies on much
larger systems will answer this question.

In summary, Fock space localization has been shown to describe the
features of the excitation statistics of many particle systems.
We have proposed a finite size scaling hypothesis 
in which the role of system size in
the usual Anderson scaling picture is played here by the excitation energy.
The scaling hypothesis was tested for a specific tight-binding many
particle Hamiltonian and resulted in satisfactory agreement.

We are grateful to A. Kamenev, 
A. M\"uller-Groeling and D. Shepelyansky for useful 
discussions and correspondence.

\end{document}